\documentstyle[mprocl]{article}

\input epsf
\def\Journal#1#2#3#4{{#1} {\bf #2}, #3 (#4)}

\def\PLB{{\em Phys. Lett.}  B}

\def\PRD{{\em Phys. Rev.} D}

\def\be{\begin{equation}}
\def\ee{\end{equation}}
\def\bea{\begin{eqnarray}}
\def\eea{\end{eqnarray}}

\begin{document}

\title{Beyond the Standard Model at Tevatron}

 \author{C. Pagliarone}

\address{University of Cassino \& INFN Pisa,\\
via di Biasio,43 - 03043 Cassino, Italy\\E-mail: pagliarone@fnal.gov}
\maketitle\abstracts{This article presents recent results
of searches for physics beyond the Standard Model using
the CDF and the D$\not$O detector at the Fermilab Tevatron
Collider. All results shown correspond to analysis performed
using the past 1992-1996 Fermilab Tevatron run I data
(roughly $100$ $pb^{-1}$ per each experiment).
In particular we describe recent Tevatron searches
for scalar top in the $b+\ell+$missing-$E_{T}$ channel,
for squark and gluinos using
like-sign dileptons (LS), for large extra space-time dimensions
and the search for leptoquarks and technicolor
in the missing-$E_{T}$+heavy flavour jet  events.
Tight limits on the existence of such models have been set.}

\section{Introduction}

The Standard Model (SM) represents the simplest and
most economical theory which describes jointly weak
and electromagnetic interactions. At present it
continues to survive all experimental tests at
accelerators, providing a remarkably successful
description of known phenomena; some precision
observable tests it at $10^{-3}$
levels~\cite{SM}.
In spite of that, there are plenty of aspects
that we do not understand yet and that may suggest the SM to be
most likely, a low energy effective theory of
spin-1/2 matter fermions interacting via spin-1 gauge
bosons~\cite{Altarelli}.
The physical motivations for the searches
described below come from the attempt to
check various possible extensions of the SM.

\section{Supersymmetry Searches}\label{subsec:susy}

An excellent candidate to a new theory, able to
describe physics at arbitrarily high energies,
is Supersymmetry
(SUSY)~\cite{SUSYTEO}.
SUSY is a larger space-time symmetry,
that relates bosons to fermions so that,
in the vast space of all viable physics
theories, SUSY is not simply a point.
Almost any theory can be supersymmetrized
and the large array of choices for spontaneous
SUSY breaking (SB) just enhance these possibilities.
A comprehensive  SUSY search is almost impossible:
i.e. the most general minimal supersymmetric extension
of the SM (MSSM) counts 124 truly independent parameters.
The strategy is then
to search for signals suggested by particular models
in which theoretical assumptions are also adopted
to reduce the number of free parameters to a few.
Even if we don't have direct
experimental evidences of SUSY,
there are remarkable theoretical properties that provide
ample motivation for its study.
SUSY describes electroweak data equally well as SM
but, in addition, allows the unification of the
gauge couplings constants, the unification of the
Yukawa couplings and
do not require the incredible fine tuning,
endemic to the SM Higgs sector.
Naturally, SUSY cannot be an exact symmetry
of the nature, as none of the predicted
spin 0 partners of the quarks or leptons
and none of the spin 1/2 partners
of the gauge bosons have been
observed so
far~\cite{BOER}.

\noindent
In SUSY models fermions can
couple to a sfermion and a fermion,
violating lepton ($L$) and/or baryon number ($B$).
To avoid this problem, a new quantum number,
the $\mathcal{R}$-parity ($\mathcal{R}_{P}$),
has been
introduced~\cite{DREINER}.
$\mathcal{R}$-parity is a multiplicative quantum number
defined as ${\mathcal{R}_{P}}=(-1)^{3B-L+2S}$,
where $S$ is the spin of the particle.
For the SM particles $\mathcal{R}_{P}=$ $+1$;
$\mathcal{R}_{P}=$ $-1$ for the SUSY partners.
Most searches for supersymmetric particles assume
$\mathcal{R}_{P}$ conservation.
This assumption
has deep phenomenological
consequences~\cite{VISSANI}:
SUSY particles can only be pair produced;
the Lightest Supersymmetric Particle (LSP) does exist
and it is stable and interacts very weakly with the
ordinary matter, leading to a robust
missing transverse energy signature ($\not\!\!\!E_{\rm T}$);
the LSP is a natural candidate for the dark matter.
However, SUSY does not require
$\mathcal{R}_{P}$ conservation
and viable $\mathcal{R}_{P}$
violating models ($\not\!\!\!\mathcal{R}_{P}$)
can be built by adding explicitly $B$-violating
and/or $L$-violating couplings to the SUSY Lagrangian.

\noindent
In this article we present both searches based
or not on the assumption of $\mathcal{R}_{P}$ conservation.

\vspace{0.2cm}
\subsection{Scalar top quark searches}

Search for scalar top is particularly
interesting since in many SUSY models
the top-squark  eigenstate
$\tilde{t}_{1}$ (stop)  is expected to be the
lightest
squark~\cite{GENSTOP}.
The strong Yukawa coupling between top/stop
and Higgs fields gives rise in fact to potentially
large mixing effects and mass splitting.\\
\noindent
Both the CDF and D$\not$O experiments have already
reported searches for direct stop quark pair
production: $p \bar{p} \rightarrow \tilde{t}_{1} \bar{\tilde{t}}_{1}$
with the $\tilde{t}_{1}$ decaying into $\tilde{t}_{1} \rightarrow  c
\tilde{\chi}^{0}_{1}$
or $\tilde{t}_{1} \rightarrow  b \tilde{\chi}^{\pm}_{1}$,
as well as for SUSY top decays: $t \rightarrow
\tilde{t}_{1}\tilde{\chi}^{0}$
with
$\tilde{t}_{1} \rightarrow b \chi^{\pm}_{1}$~\cite{stop1}.\\
\noindent
CDF recently searched for $\tilde{t}_{1} \bar{\tilde{t}}_1$
production, assuming $\mathcal{R}_{P}$ conservation,
within the framework of the MSSM, for the case where
$m_{\tilde{t}_1}< m_{t}$.
Two separate  $\tilde{t}_{1}$ decay modes have been considered:
$\tilde{t}_{1} \rightarrow b\tilde{\chi}^{+}_{1}$,
with the chargino decaying into
$\ell^{+}\nu \tilde{\chi}^{0}_{1}$ ($\ell= e, \mu$)
and $\tilde{t}_{1} \rightarrow b\ell^{+}\tilde{\nu}$,
dominant whenever  $\tilde{t}_{1} \rightarrow b\tilde{\chi}^{+}_{1}$
is not kinematically allowed.
The search have been done
assuming the following branching ratios for the involved processes:
$\mathcal{B}$$(\tilde{t}_{1} \rightarrow b\tilde{\chi}^{+}_{1})= 100\%$,
${\mathcal{B}}(\tilde{\chi}^{+}_{1} \rightarrow \ell^{+}\nu
\tilde{\chi}^{0}_{1})= 11\%$
and ${\mathcal{B}}(\tilde{t}_{1} \rightarrow
b\ell^{+}\tilde{\nu})=33.3\%$.
In these two scenarios, either the  $\tilde{\chi}^{0}_{1}$ or the
$\tilde{\nu}$ is the LSP.
The event signature is 2$b$-jets+$\ell$+ $\not\!\!\!E_{\rm T}$
in both the cases;
hence events have been selected requiring the presence of at least
one $e$ or $\mu$ with $p_{T}^{\ell}> 10$ $GeV/c$, two or more
jets, with cone size of $R=\sqrt{\Delta\eta^{2}+\Delta\phi^{2}}= 0.7$
and $E_{T}^{{jet}_{1}}> 12$ $GeV$, $E_{T}^{{jet}_{2}}> 8$ $GeV$
(jets have been ordered by energy) and large missing-$E_{T}$ from the
neutral LSP's:
$\not\!\!\!E_{\rm T}> 25$ $GeV$ with
$\Delta\phi(\not\!\!\!E_{\rm T},jet)> 0.5$~\cite{cdf_cord}.
At least one jet has to be identified as a
$b$-jet candidate using a method known as SVX
tagging~\cite{cdf_top}.
The significant SM backgrounds:
$t \bar{t}$,  $b \bar{b}$, $W^{\pm}$$(\rightarrow \ell^{\pm}\nu)+2$ jets
and
fake lepton events
are predicted to contribute for $87.3\pm8.8$ events
while $81$ are observed in the data.
To determine the number of potential signal events in this final data
sample for both decay scenarios we used
an extended unbinned likelihood fits for each
$\tilde{t}_{1}$ mass considered.
The likelihood fits compare the shapes of the signal
and backgrounds distributions, and Kolmogorov tests are
used in order to determine the most sensitive kinematic distribution
to use in the fit.
All fit results at all masses are consistent with zero signal events.
The 95\%
{\it C.L.} limits on $\sigma_{\tilde{t}_{1}\bar{\tilde{t}}_1}$
for $\tilde{t}_{1} \rightarrow b \tilde{\chi}^{+}_{1}$
\,and \,for  \,$\tilde{t}_{1} \rightarrow b \ell^{+} \tilde{\nu}$
\,as \,function \,of \,$m_{\tilde{t}_1}$ \,are
\newpage
\begin{center}
\begin{figure}[ht!]
\begin{minipage}{5.0in}
  \epsfxsize2.75in
  \vspace*{-2.0cm}\hspace*{-0.4cm}\epsffile{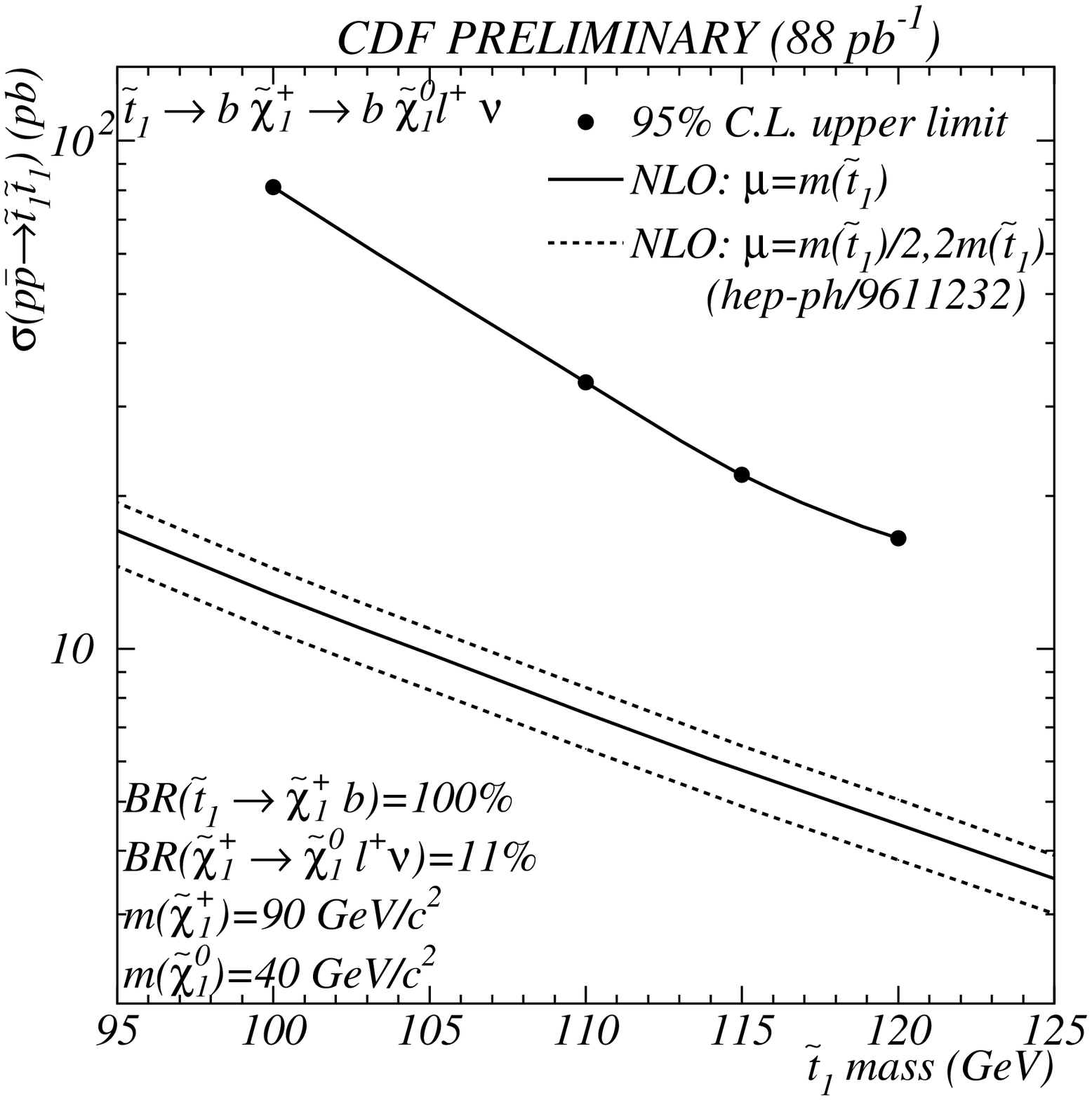}
  \epsfxsize2.75in
  \vspace*{-0.6cm}\hspace*{-0.55cm}\epsffile{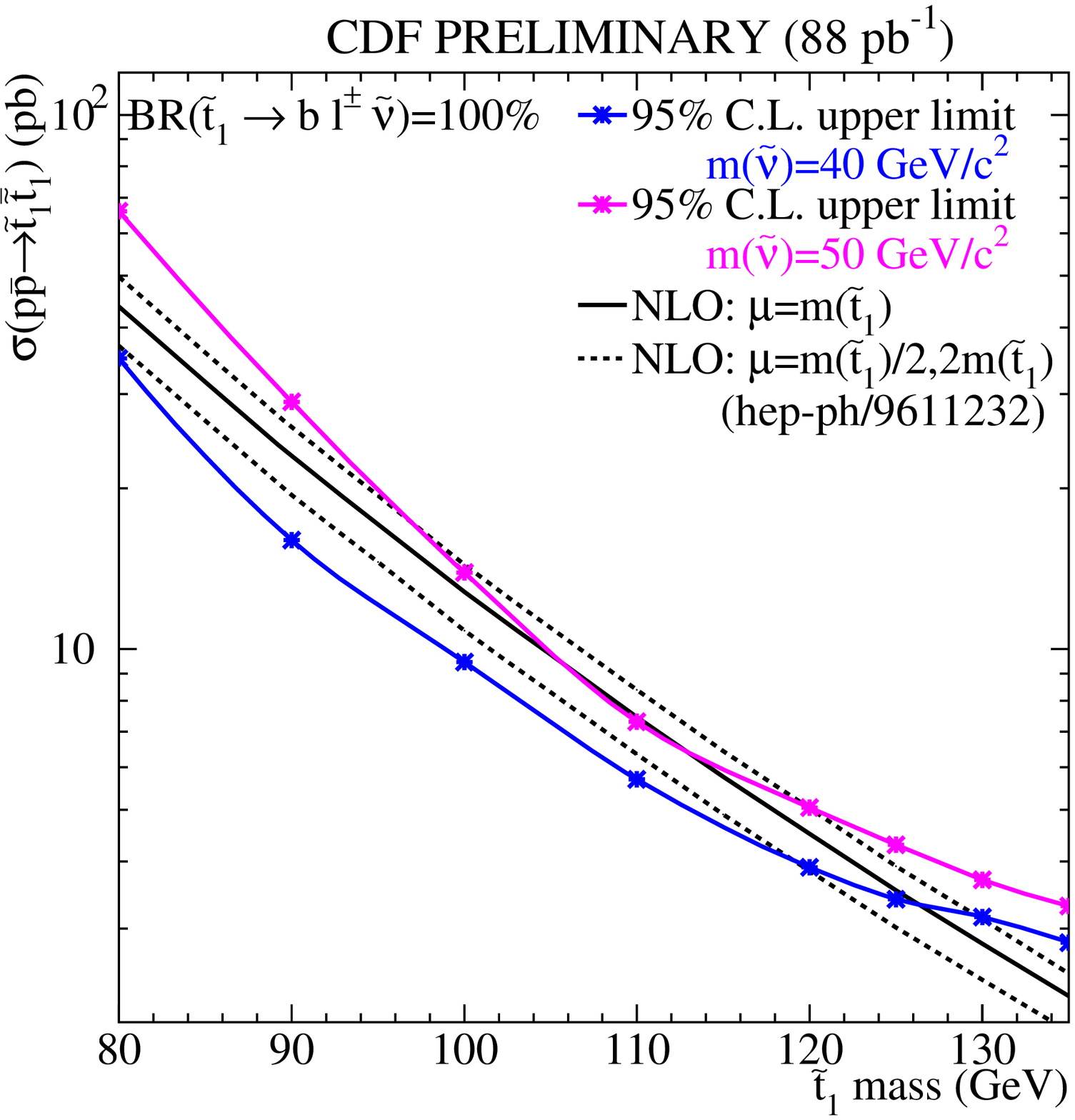}
\end{minipage}\hfill
\end{figure}
\begin{figure}[th!]
\vspace{-0.0cm}
\begin{minipage}{2.4in}
\vspace{-0.0cm}
  \caption{\it CDF 95\% C.L. Cross section limit as function of
$\tilde{t}_{1}$,
assuming a branching ratio
${\mathcal{B}}(\tilde{t}_{1}\rightarrow b \chi^{+}_{1})=100\%$ and
$m_{\tilde{\chi}^{\pm}_{1}}= 90$ $GeV/c^{2}$,
$m_{\tilde{\chi}^{0}_{1}}= 40$ $GeV/c^{2}$.}
\label{stop_1a}
\end{minipage}\hfill
\hspace{0.2cm}
\begin{minipage}{2.4in}
\vspace{-0.19cm}
  \caption{\it  CDF 95\% C.L. Cross section limit as function of
$\tilde{t}_{1}$, assuming a branching ratio
${\mathcal{B}}(\tilde{t}_{1}\rightarrow b \ell^{+}\tilde{\nu})=100\%$
and $m_{\tilde{\nu}}= 40$ $GeV/c^{2}$ or $50$ $GeV/c^{2}$.}
\label{stop_1b}
\end{minipage}\hfill
\vspace{-.39cm}
\end{figure}
\end{center}
\vspace{0.2cm}
\noindent
shown in
Figure~\ref{stop_1a}
and~\ref{stop_1b}.
The resulting excluded regions
($m_{\tilde{t}_{1}}$ versus $m_{\tilde{\nu}}$), for
$\tilde{t}_{1} \rightarrow b \ell^{+} \tilde{\nu}$
are given in
Figure~\ref{stop_2}.

\vspace{0.5cm}
\subsection{Search for squarks and gluinos using LS dileptons}

   Gluinos ($\tilde{g}$) and squarks ($\tilde{q}$)
can decay, via charginos  ($\chi^{\pm}$)
and neutralinos ($\chi^{0}$),
to final states containing two o more leptons. Since the gluino
is a Majorana particle, a large fraction of like-sign dilepton (LS)
events will be produced.
Squarks and gluinos have been searched by Tevatron experiments
looking mainly for $\not\!\!\!E_{\rm T}$+multijet
events~\cite{gluinoclassic}.
Complementary to these classic $\not\!\!\!E_{\rm T}$+multijet
searches,
recently CDF looked for gluinos decaying through
charginos or neutralinos
giving the LS
signature: $\ell^{\pm}\ell^{\pm}$$+jets+$$\not\!\!\!E_{\rm T}$.\\
\noindent
Data have been selected by asking for an isolated central lepton
($e$ or $\mu$) with $E^{\ell_{1}}_{T}>$ $11$ $GeV$
and for a second one with $E^{\ell_{2}}_{T}>$ $5$ $GeV$.
In order to reject background from $B$'s,
leptons were asked to be well separated in the $\eta$-$\phi$ plane.
Events were also required to contain two jets in the pseudorapidity
region $|\eta|$$<$ $2.4$ with $E_{T}> 15$ $GeV$ and
at least 25 GeV of $\,\,\not\!\!\!E_{\rm T}$.
Then the like-sign requirement
on the dilepton event sample have been applied.\\
\noindent
No candidate events have been observed, which is consistent
with the SM background expectations
(mostly $t\bar{t}$, $b\bar{b}$, diboson, and Drell-Yan processes).
The results of the analysis are shown in
Figure~\ref{jpdone}
where the 95\% {\it C.L.}
exclusion region is plotted as function of squark versus gluino mass.
The model assumes tan$\,\beta$= $2$, $\mu$$=$ $-800$ $GeV/c^2$,
$A_{t}=$ $\mu$/tan$\,\beta$,
$A_{b}=$ $A_{\tau}=$ $\mu$ tan$\,\beta$, and
$m_{A^{0}}=$ $500$ $GeV/c^2$.
Constraining the squark masses to have an infinite value,
we can exclude gluino masses up to $168$ $GeV/c^2$;
assuming the squark and sgluinos masses nearly equal
then we reach a limit of $221$ $GeV/c^2$.

\newpage
\begin{center}
\begin{figure}[t]
\begin{minipage}{8.0in}
  \epsfxsize2.75in
  \vspace*{0.0cm}\hspace*{-0.55cm}\epsffile{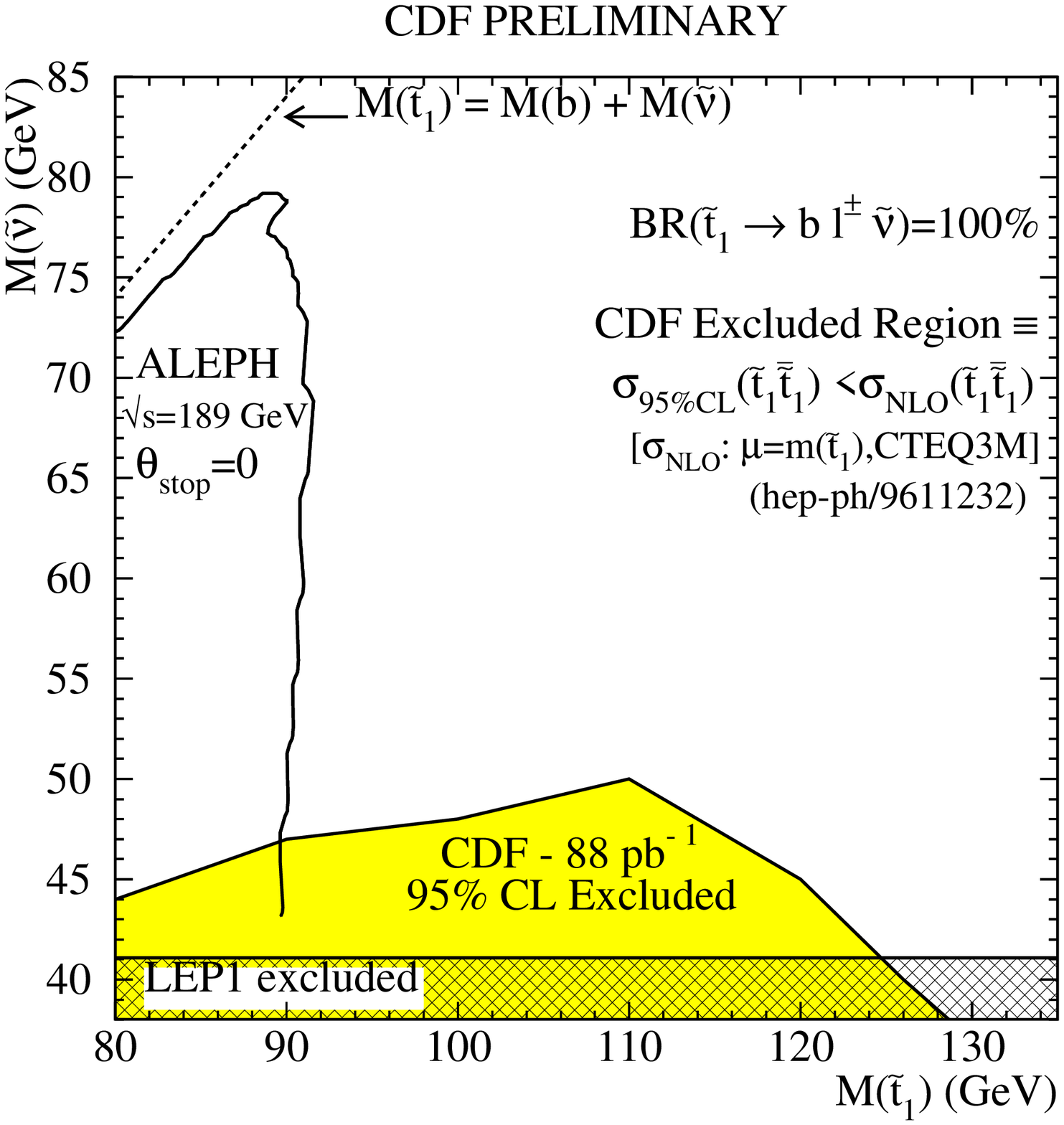}
  \epsfxsize2.5in
  \epsfysize2.5in
  \vspace*{0.0cm}\hspace*{0.15cm}\epsffile{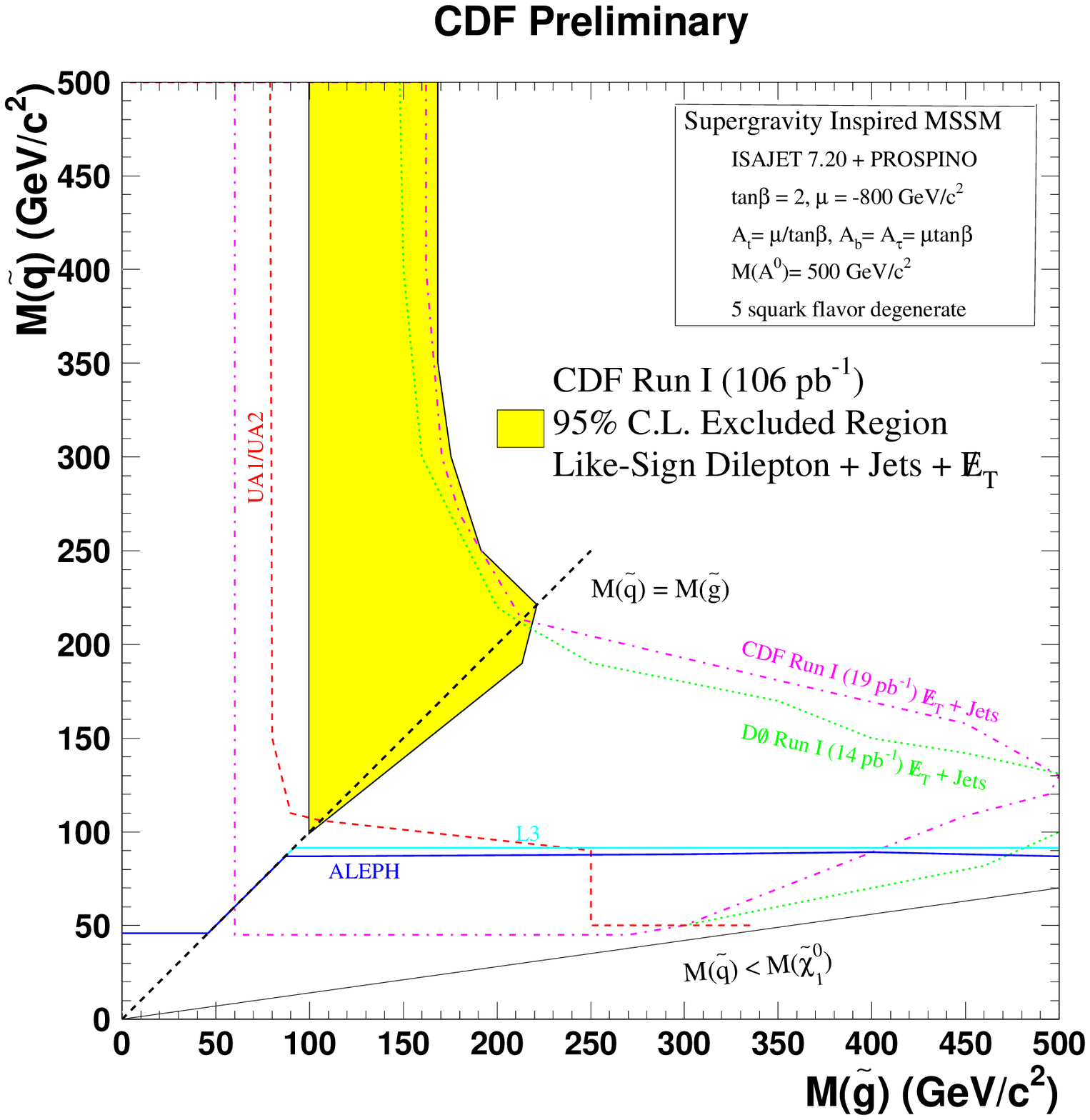}
\end{minipage}\hfill
\end{figure}
\begin{figure}[h!]
\vspace{-0.7cm}
\begin{minipage}{2.4in}
\vspace{-0.0cm}
  \caption{\it CDF 95\% C.L. excluded region in the
               $m_{{\tilde{t}}_{1}}$ versus $m_{\tilde{\nu}}$
for ${\mathcal{B}}(\tilde{t}_{1}\rightarrow b
\ell^{+}\tilde{\nu})=33.3\%$
($\ell=e\,\mu\,\tau$).}
\label{stop_2}
\end{minipage}\hfill
\hspace{0.2cm}
\begin{minipage}{2.4in}
\vspace{-0.11cm}
  \caption{\it CDF Search for $\tilde{g}$ decaying through $\tilde{\chi}^{\pm}$ or
               $\tilde{\chi}^{0}$ giving like-sign dileptons events; 95\% CL excluded 
	       region.}
\label{jpdone}
\end{minipage}\hfill
\vspace{-.39cm}
\end{figure}
\end{center}

\subsection{Search for $\mathcal{R}$-parity violation in multilepton
channel}

\noindent
There exists a well-motivated Minimal SUSY extension of the SM
in which supersymmetry breaking is communicated to the SM
particles and to the sparticles via gravitational
interactions (mSUGRA).
A mSUGRA model with $\mathcal{R}_{P}$ conservation
has only $5$ free parameters.
Allowing  $\mathcal{R}_{P}$ non-conservation
requires $45$ more parameters.
Then general super-potential can be written as:
$W_{MSSM}+\lambda_{ijk}L_{i}L_{j}\bar{E}_{k}+
\lambda^{'}_{ijk}L_{i}Q_{j}\bar{D}_{k}+
\lambda^{''}_{ijk}\bar{U}_{i}\bar{D}_{j}\bar{D}_{k}+
\mu_{i}H_{2}L_{i}$
where we have $45$ Yukawa-type coupling terms
($9$ $\lambda_{ijk}$, $27$
$\lambda_{ijk}^{'}$ and 9 $\lambda_{ijk}^{''}$).
If the $B$-violating terms ($\lambda^{''}$) lead to
events with multijet signatures, difficult to study at hadron colliders,
$L$-violating terms ($\lambda$ and $\lambda^{'}$)
give rise to multilepton and multijet final states.
Recently  D$\not$O searched for mSUGRA with
$\not\!\!\!\mathcal{R}_{P}$
assuming non-zero $\lambda_{ijk}$ couplings
then allowing the decay of LSP
(assumed to be $\tilde{\chi}^{0}$)
into two charged leptons and a neutrino.
The model also assumes that only one $\lambda_{ijk}$
is non-zero at a time and that
$0.001<\lambda<0.01$
that means that $\not\!\!\!\mathcal{R}_{P}$
introduces negligible changes in the
production and decay of SUSY particles and
that LSP is forced to decay at less than 1 cm
from the vertex ($\lambda> 0.001$).
The event signature is then 2 LSPs that decay, yelding events with
$\ge 4 \ell + \not\!\!\!E_{\rm T}$.
This search reinterpreted the results of a previous
D$\not$O search for gaugino pair production
($\tilde{\chi}^{\pm}_{1}\tilde{\chi}^{0}_{2}$)
in multilepton
channel~\cite{RPVbiblio}.
The event selection and background estimations used
are the same of this work; trilepton events
($\ell=e,\mu$) are selected requiring
$0<|\eta(e)|<1.2$, $1.4<|\eta(e)|<3.5$ and
$0<|\eta(\mu)|<1.2$ and by applying cuts on
$E_{T}^{\ell}$, $\Delta\phi(\ell\ell)$, $\not\!\!\!E_{\rm T}$
and  $\Delta\phi(\mu,\not\!\!\!E_{\rm T})$
to remove instrumental background and cosmic rays.
No candidate events have been observed,
which is consistent with the expected
SM background (see table~\ref{table1}).

\newpage
\begin{figure}[t!]
\hspace{-1.5cm}
\vspace{-0.3cm}
\begin{minipage}{2.0in}
  \epsfxsize3.2in
  \hspace*{-0.2cm}\epsffile{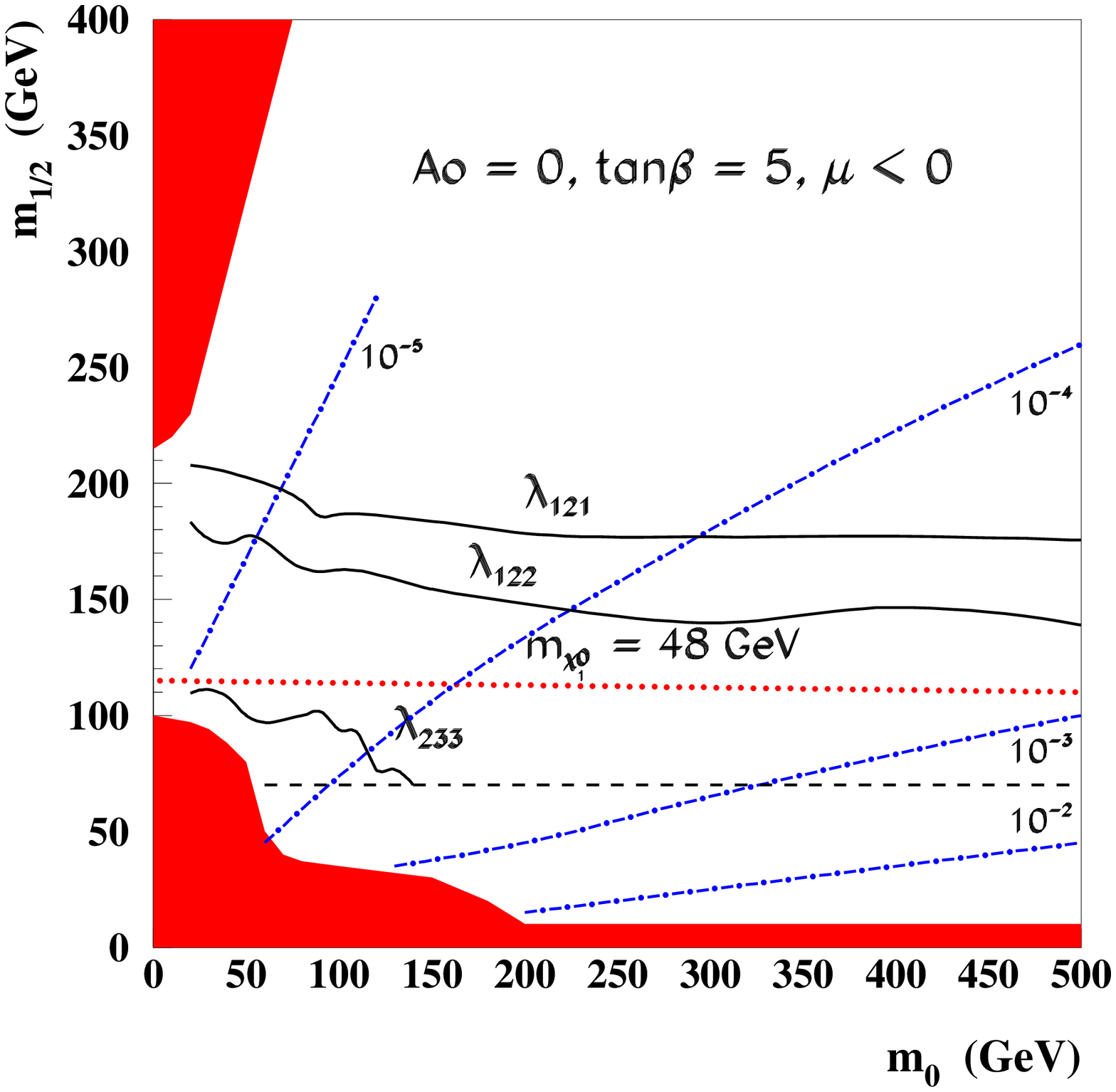}
  \epsfxsize3.2in
  \hspace*{-0.5cm}\epsffile{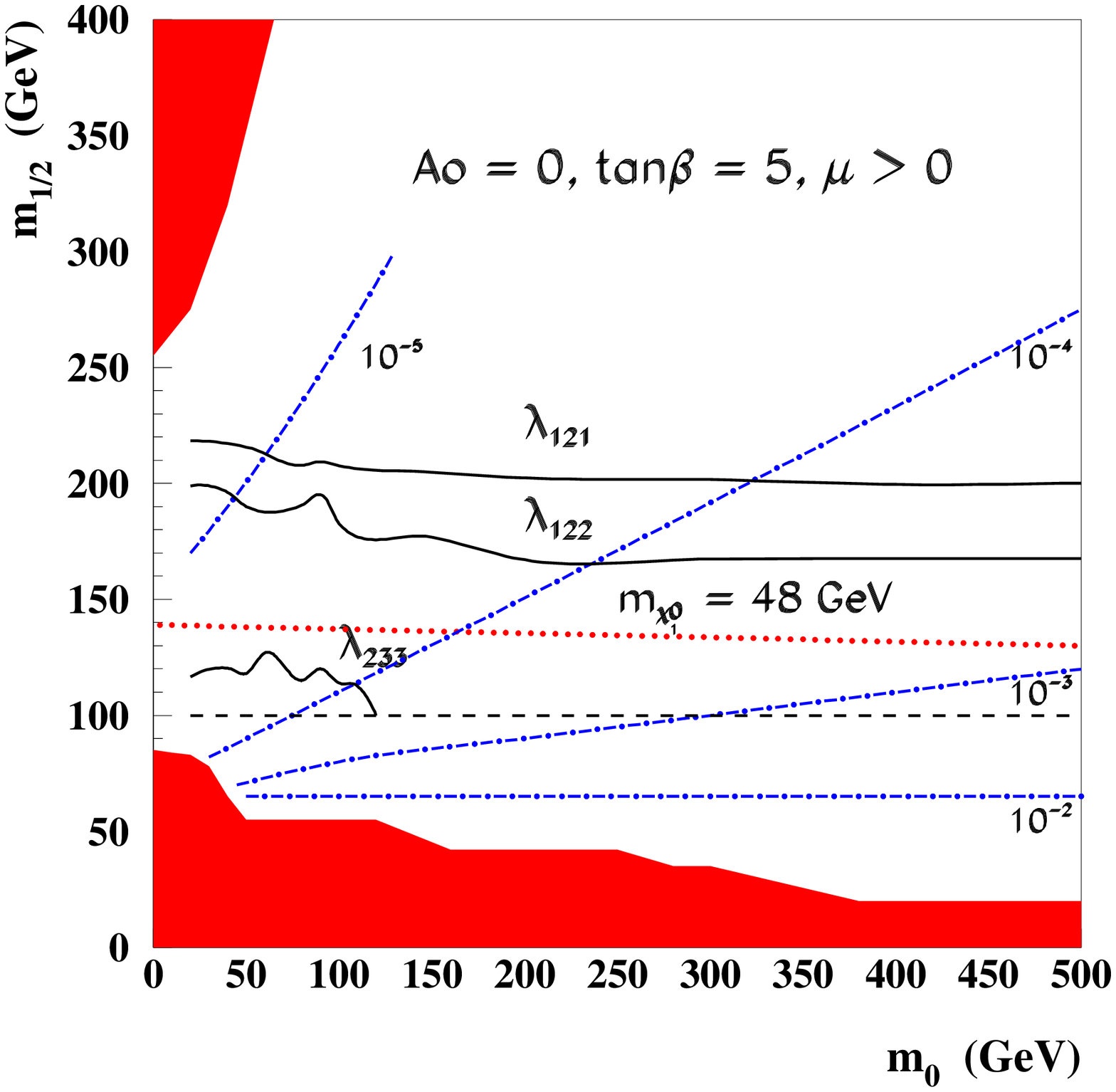}
\end{minipage}\hfill
\caption{\it  D$\not$O Search for $\not\!\!\!\mathcal{R}_{P}$
SUSY in multilepton events. The plots show
results in the mSUGRA mass plane for $A^{0}=0$, tan $\beta$= $5$
and $\mu<$ $0$ (left) or $\mu>$ $0$ (right).}
\label{RPV1}
\end{figure}
\vspace{-1.5cm}
\begin{figure}[h!]
\hspace{-1.5cm}
\vspace{-0.3cm}
\begin{minipage}{2.0in}
  \epsfxsize3.2in
  \hspace*{-0.2cm}\epsffile{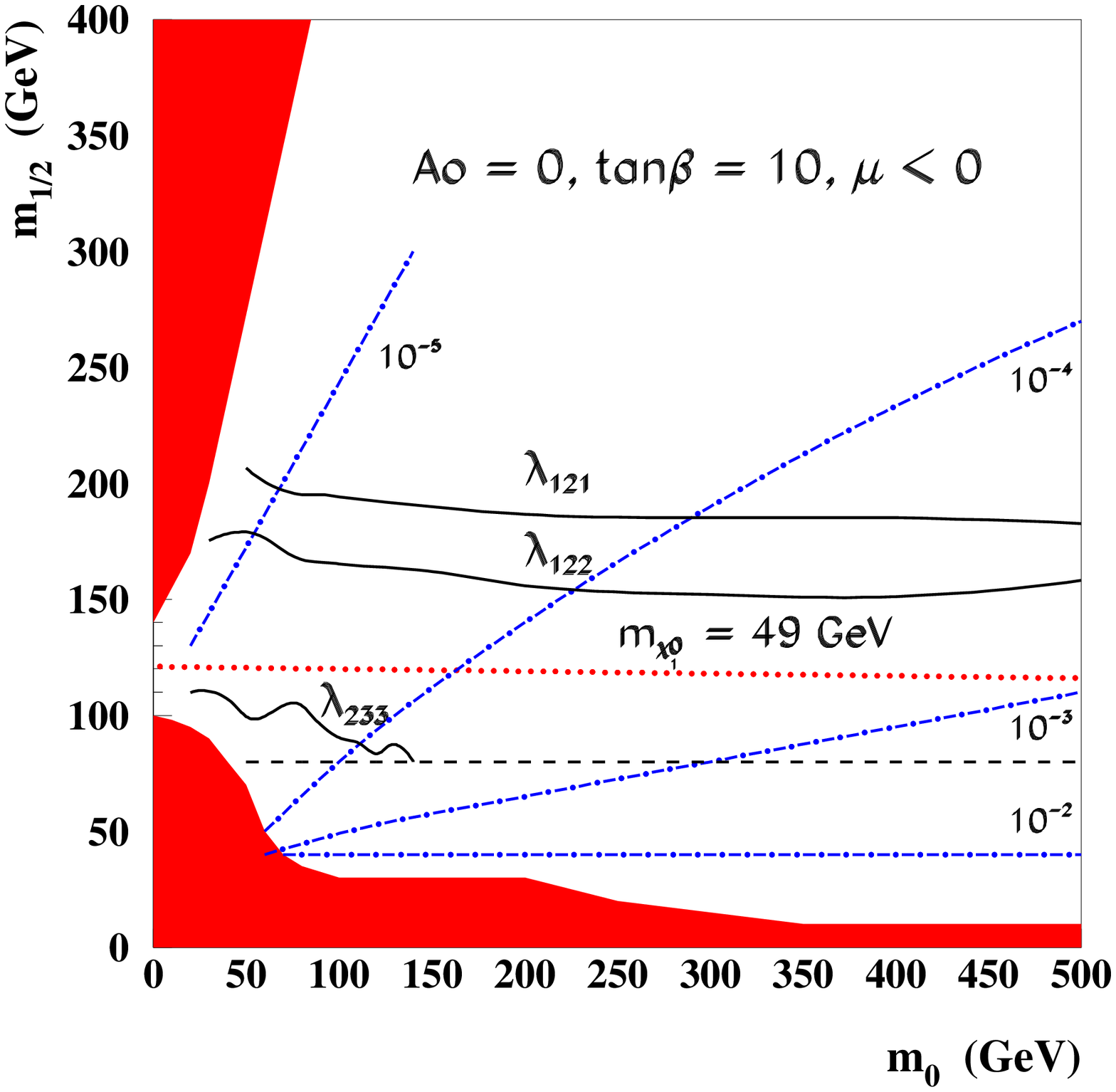}
  \epsfxsize3.2in
  \hspace*{-0.5cm}\epsffile{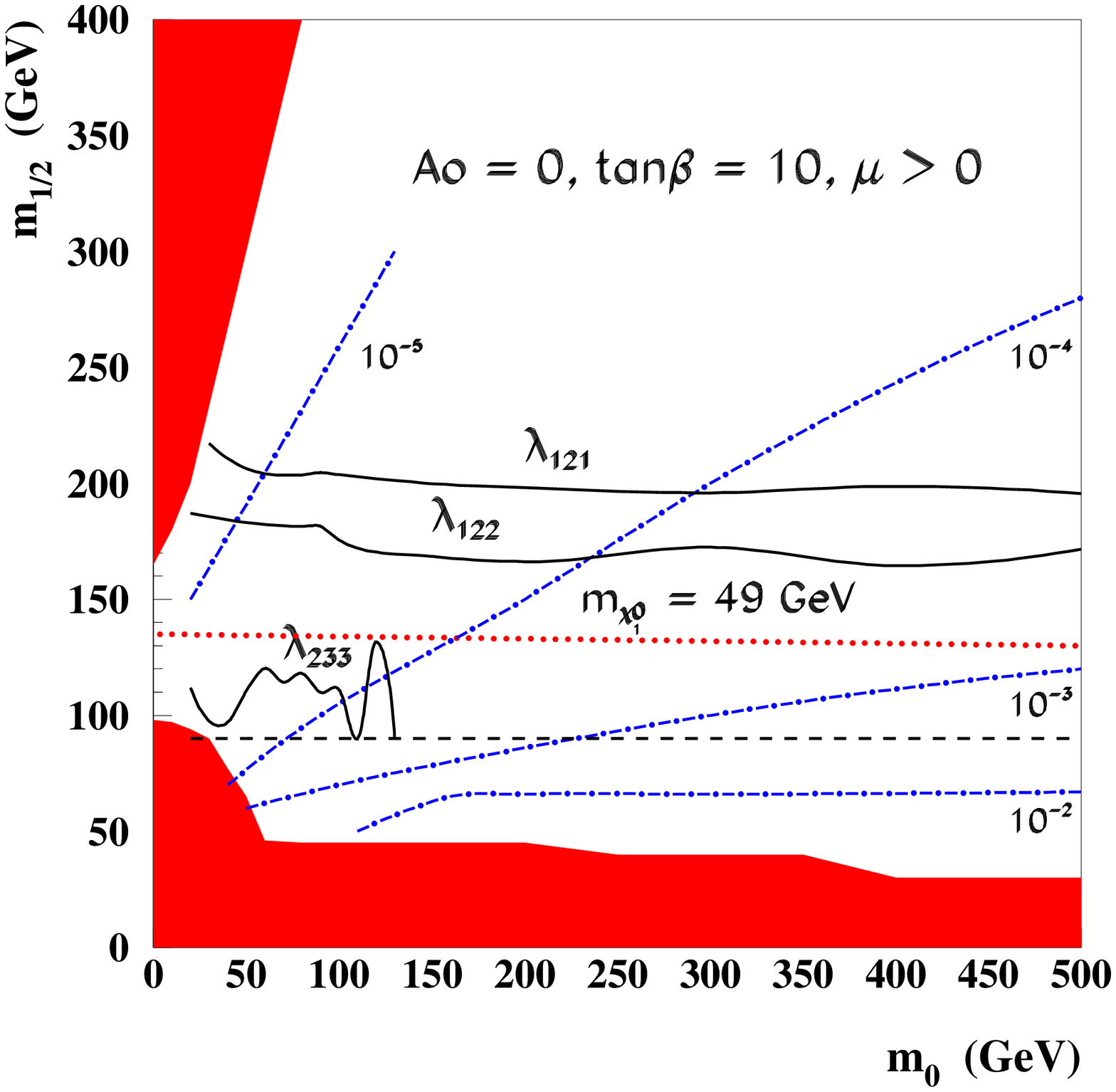}
\end{minipage}\hfill
\caption{\it  D$\not$O Search for $\not\!\!\!\mathcal{R}_{P}$
SUSY in multilepton events. The plots show
results in the mSUGRA mass plane for $A^{0}=0$, tan $\beta$= $10$
and  $\mu<$ $0$ (left) or $\mu>$ $0$ (right).}
\label{RPV2}
\end{figure}

\newpage
\begin{table}[t!]
\centering
\begin{tabular}{|l|c|c|c|c|}  \hline\hline
Event categories           & $eee$          & $ee\mu$       &    $e\mu\mu$   & $\mu\mu\mu$   \\
\hline\hline
$\mathcal{L}$($pb^{-1}$)   & 98.7$\pm$5.2   &  98.7$\pm$5.2 &   93.1$\pm$4.9 &  78.3$\pm$4.1 \\
\hline
Observed events            & 0              &  0            &   0            &  0            \\
\hline
Background events          & 0.34$\pm$0.007 & 0.61$\pm$0.36 &  0.11$\pm$0.04 & 0.20$\pm$0.04 \\
\hline\hline
\end{tabular}
\caption{\it  D$\not$0: Results of the search for trilepton events.}
\label{table1}
\end{table}

\noindent
Limits on $\not\!\!\!\mathcal{R}_{P}$
mSUGRA models can be set as a function of the following $5$
parameters: $m_{0}$, $m_{1/2}$, $A_{0}$, tan $\beta$ and $\mu$.
Figures~\ref{RPV1} and~\ref{RPV2} show, respectively, the exclusion
regions in the ($m_{0}$, $m_{1/2}$) plane for the three chosen couplings, for
tan $\beta=5$ and tan $\beta=10$ for both  $\mu>0$ and  $\mu<0$.
The dashed lines indicates the limit of sensitivity in $m_{1/2}$
for at least favourable case. The exclusion regions correspond to the
spaces below the solid lines labelled with the coupling types, and
above the higher of the dashed line and the dash-dotted curves
specifying the numerical values of $\lambda$.
In the regions beyond the dash-dotted curves, the
average decay length of the LSP calculated for the value of the
coupling indicated on the curve, is less than 1 cm.


\section{Searches for Large Extra Dimensions}

\begin{figure}[t!]
\hspace{-0.8cm}
\begin{minipage}{2.5in}
  \epsfxsize2.75in
  \hspace*{0.0cm}\epsffile{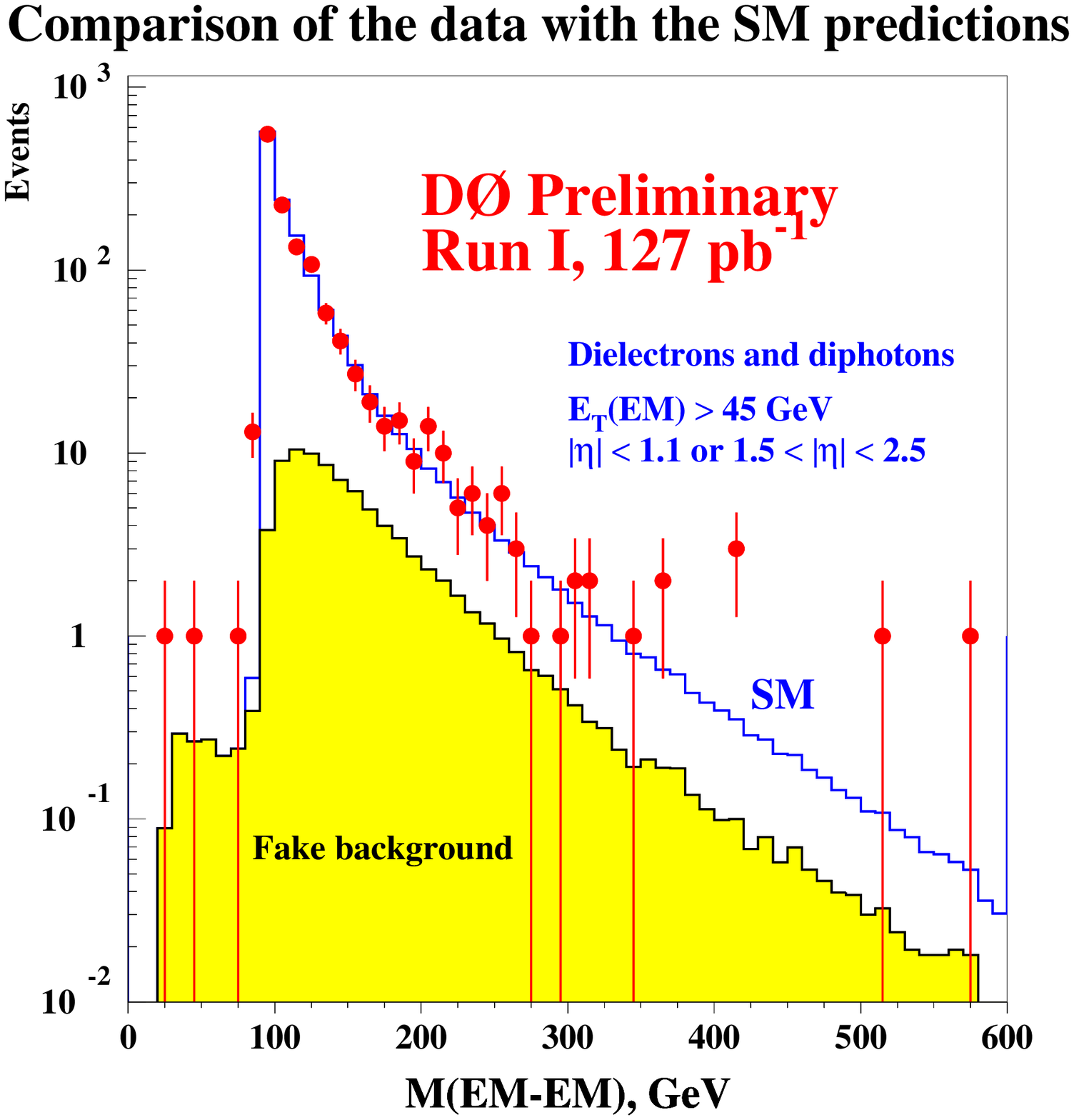}
  \epsfxsize2.75in
  \hspace*{-0.1cm}\epsffile{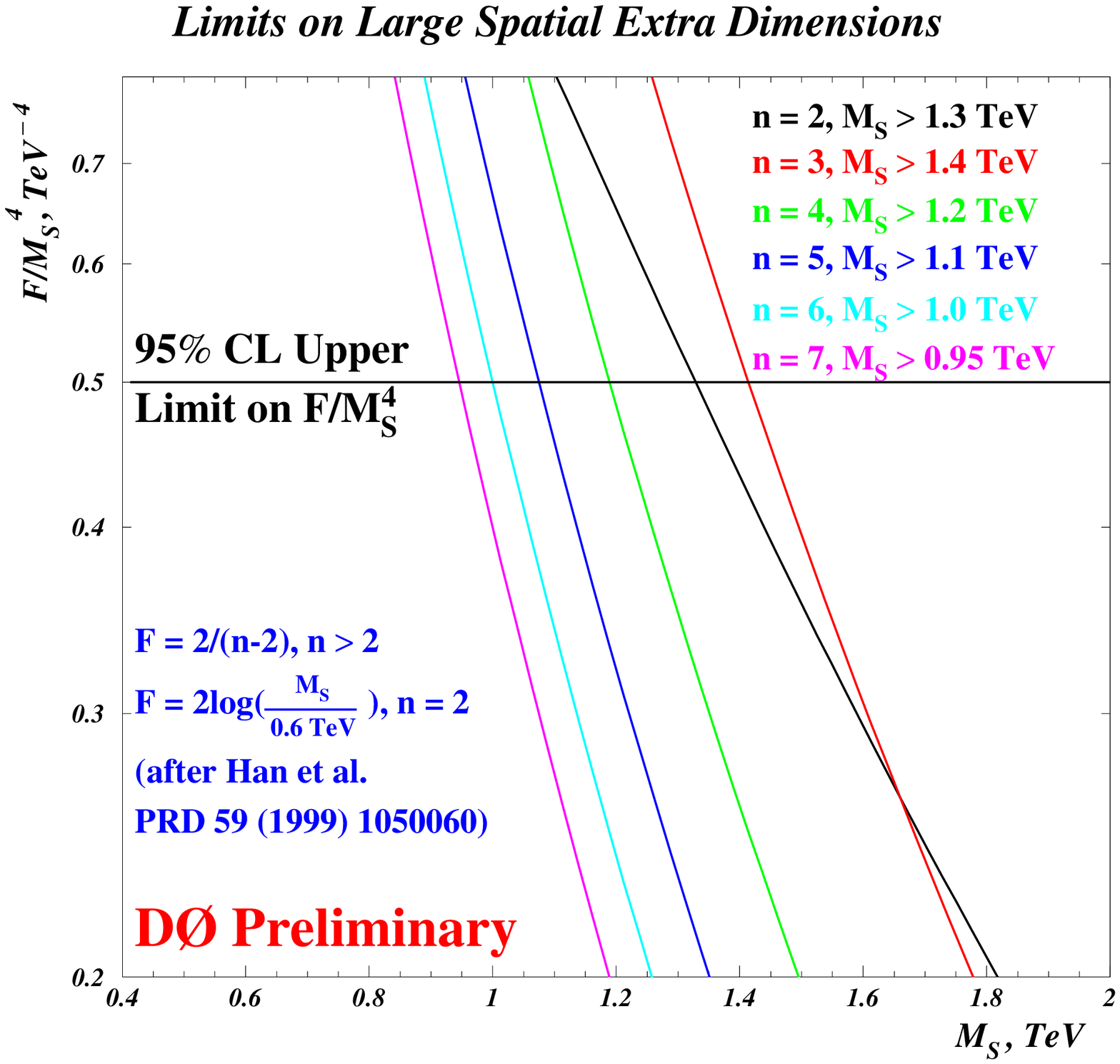}
\end{minipage}\hfill
\caption{\it  D$\not$O Dielectron and diphoton invariant
mass distributions compared to the expected SM Drell-Yan
and instrumental background contributions (left);
limits on large extra dimensions from the dielectron and
diphoton (right).}
\label{ExtraDim}
\end{figure}

   Motivated in part by naturalness issues, recently,
it has been pointed out that there are more
reasons to suggest extra dimensions than just having a
self-consistent description of
gravity~\cite{Arkani1}.
The additional motivations include new directions to attack
the hierarchy problem and the cosmological constant
problem, unifying the gravitational coupling with
the gauge couplings, perturbative supersymmetry
breaking in string theory, and low-scale compactifications
of string theory.
In such scenario, the hierarchy problem could be solved by
having gravity play a role at the electroweak scale.
The SM fields are then confined to propagate on a $3+1$ $d$-brane,
and the gravity lives in the bulk of the $(4+n)$
dimensional space-time, where the $n$ dimensions are compactified~\cite{Arkani2}.
The size of extra dimensions is
$M^{2}_{Pl}$ $\sim$ $M^{n+2}_{S}R^{2}$
where $M_{S}$, the compactification-scale, is the only fundamental
scale of the nature. As $M_{S}$ could lie in the TeV range
this would offer the possibility that effects
might be visible at Tevatron Collider.
In particular indirect effects of massive graviton exchange
may be enhanced by the sum of the Kaluza-Klein
states ($KK$) and provide various signals in the collider
phenomenologies. Moreover the spin-two nature of gravitons
will also result in some characteristic effects on the
polarisation observables.
In the absence of evidence for extra dimensions
is becoming common to set 95\% $CL$ limits as
function of $\mathcal{F}$$/M^{4}_{S}$,
where
$\mathcal{F}$ is a dimensionless parameter related to the
number of extra dimensions.\\
\noindent
D$\not$O has recently carried out a search
on the effect of $KK$ tower exchange on
Drell-Yan dielectron and diphoton production,
fitting $127$ $pb^{-1}$ of data to 2-dimensional
templates in $M_{\ell\ell}$ versus $cos$$\theta^{*}$.
No evidence for these effects has been found and
lower limit on the parameter $\mathcal{F}$$/M^{4}_{S}$
as function of the number of extra dimensions
have been obtained.
For $n$ values between $2$ and $7$, $M_{S}$ values below
$1.3-1.0$ are respectively ruled out (see
Figure~\ref{ExtraDim}).

\newpage

\noindent
Analogous searches for extra dimension in the Drell-Yan, diphoton and
$\not\!\!\!E_{\rm T}$ channels are in progress at CDF.

\section{Search for Leptoquark and Technicolor}

    The observed symmetry, in the spectrum of fundamental
particles, between quarks and leptons suggests that,
if there exists a more fundamental theory, it should
also introduce a more fundamental relation between them.
Such lepton-quark unification is achieved in the context of
different theories. Leptons and quarks may be arranged in
common multiplets, like in Grand Unified Theories (GUT) or
superstring motivated $E_{6}$ models or they may have a common
substructure as in composite
models~\cite{leptoquarks}.
Whenever quarks and leptons are allowed to couple directly
to each other, a quark-lepton bound state can also exist.
Such particles, called {\it leptoquarks},
are color triplet bosons ($SU(3)_{C}$) of spin 0 or 1,
carrying lepton ($L$) and baryon ($B$) number and a
fractional electric charge.
At Tevatron, leptoquarks are predicted to be produced
dominantly via gluon-gluon fusion and $q$$\bar q$ annihilation:
$p \bar p \rightarrow g+X$ $\rightarrow$ $LQ {\bar{L}\bar{Q}}+X$.
The production cross section is nearly independent from the Yukawa
couplings between the $LQ$s and their decay lepton-quark pairs.\\
\noindent
Both CDF and D$\not$O performed in the past years a number of searches
for
leptoquarks~\cite{acosta}.
In this paper we report the most recent Tevatron results.\\
\noindent
In contrast to the previous searches,
sensitive to branching ratios
$\mathcal{B}$$(LQ \rightarrow \ell+jets)>0$,
CDF has searched for second and third generation leptoquarks
in two different theoretical context:
a) continuum $LQ$s pair production with
$LQ2 \rightarrow e\nu_{\mu}$ and $LQ3 \rightarrow b\nu_{\tau}$
($LQ$s are either scalar or vector);
b) resonant Technicolor $LQ$s.

\noindent
Final states which contain two heavy flavor jets
($c$-jets or $b$-jets for second or third

\newpage
\begin{figure}[t!]
\hspace{-0.8cm}
\begin{minipage}{2.5in}
  \epsfxsize2.75in
  \hspace*{0.0cm}\epsffile{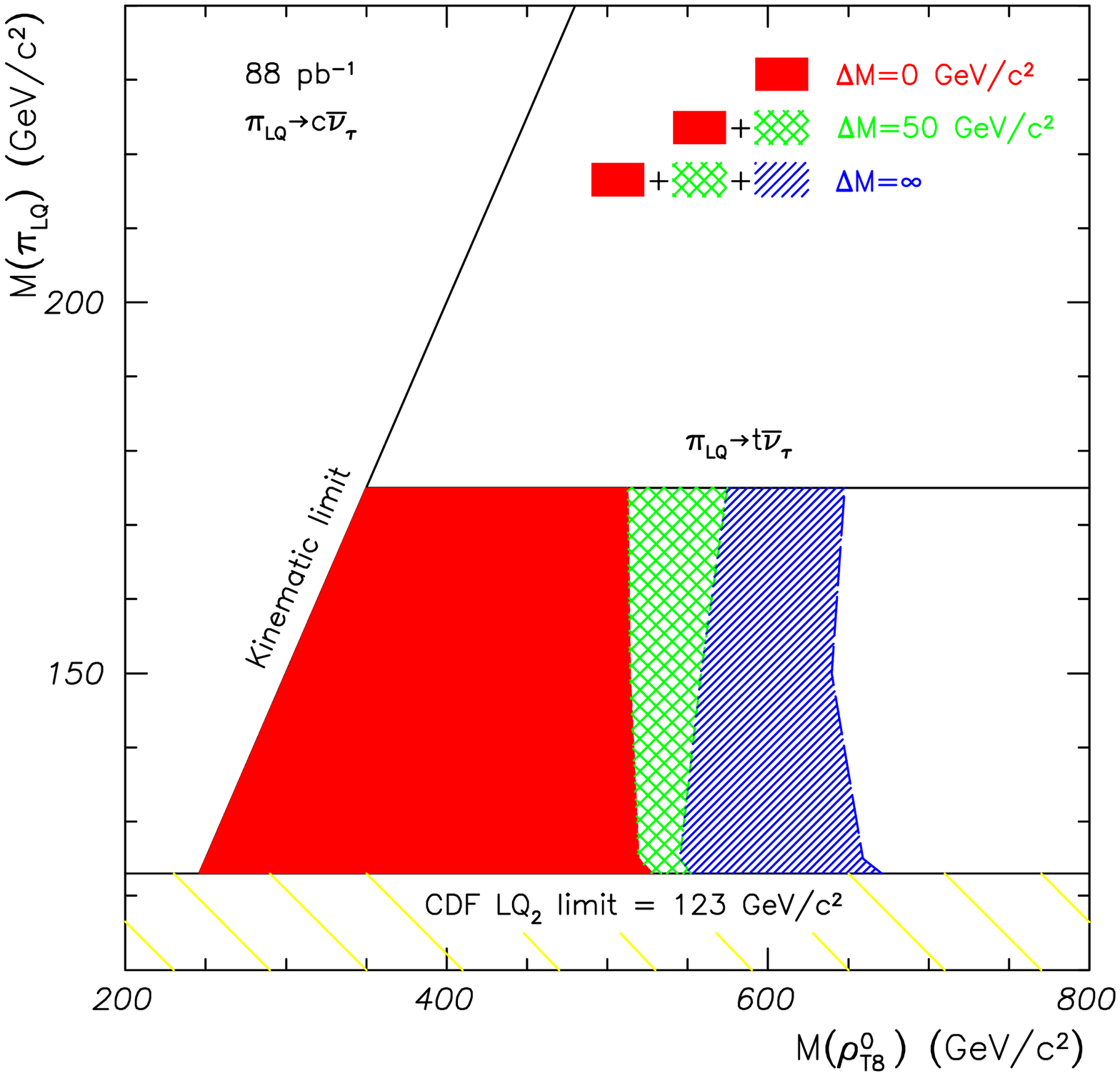}
  \epsfxsize2.75in
  \hspace*{-0.1cm}\epsffile{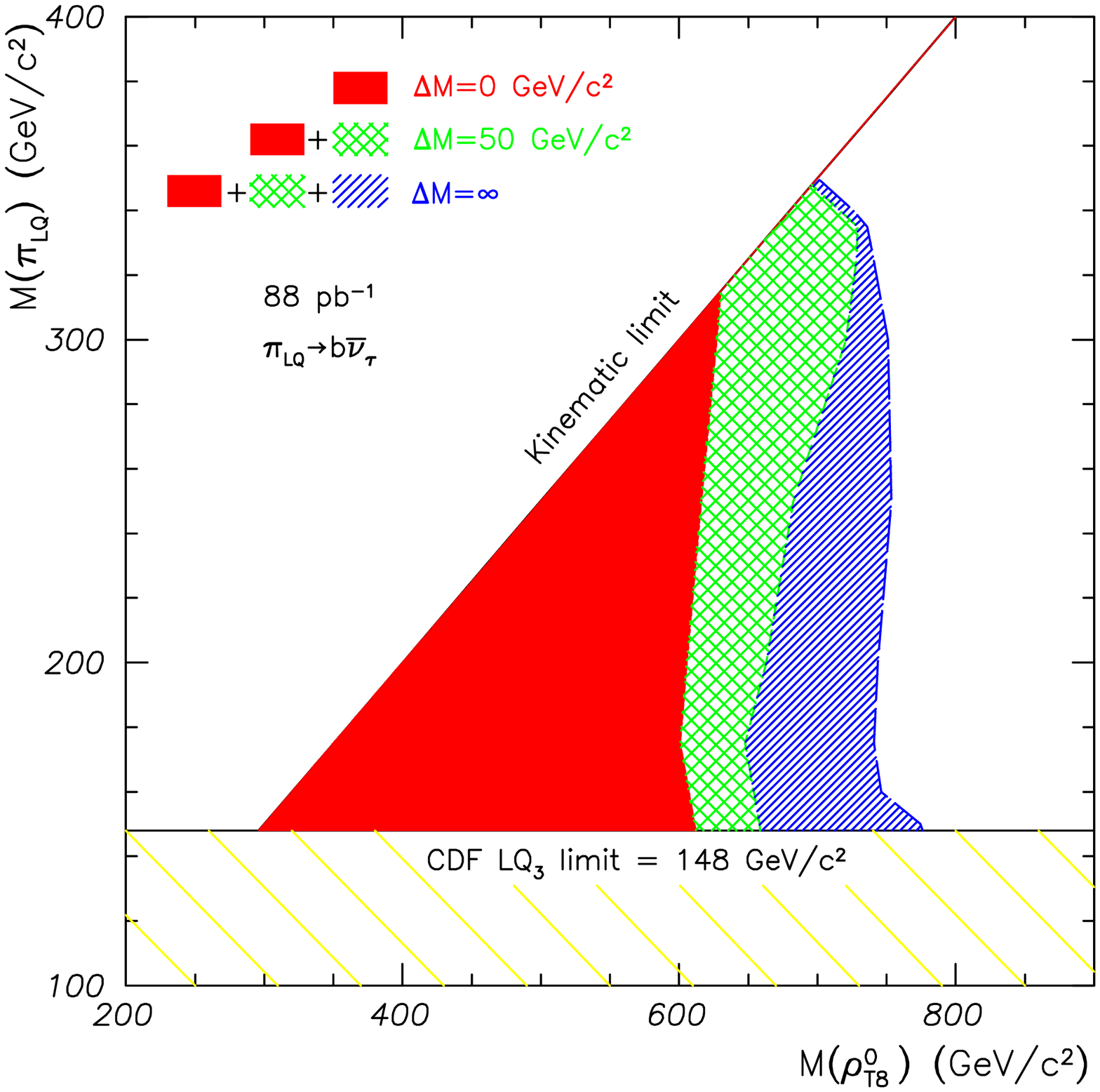}
\end{minipage}\hfill
\caption{\it CDF Search for leptoquarks in $\not\!\!\!E_{\rm T}$
plus hevy flavour jet events: 95\% C.L. limits on $\pi_{LQ}$ producion
via color-octet technirho decays.}
\label{Leptoquark1}
\end{figure}
\begin{center}
\begin{figure}[t!]
  \epsfxsize3.8in
\vspace*{-0.8cm}
\hspace*{1.7cm}\epsffile{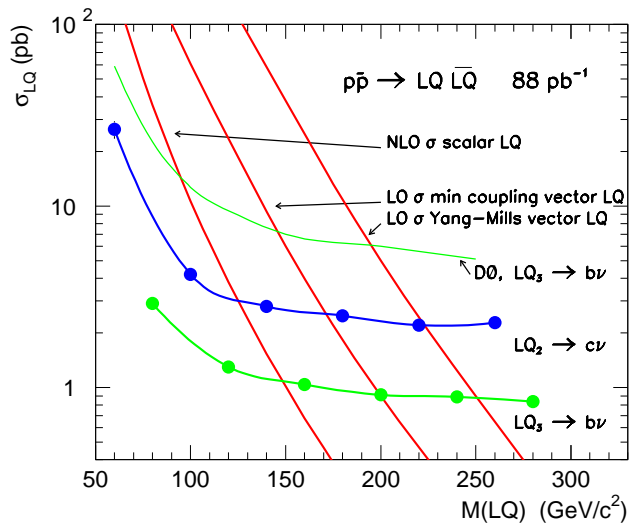}
\vspace*{-0.4cm}
\caption{\it CDF Comparison between the 95\% C.L. upper limit
on the production cross section
for $LQ_{2}\rightarrow\nu_{\mu}c$ and
$LQ_{3}\rightarrow\nu_{\tau}b$ and the theoretical
predictions.}
\label{Leptoquark2}
\end{figure}
\end{center}


\noindent
generations of $LQ$s), large missing-$E_{T}$ (from neutrinos),
and no high-$p_{T}$ leptons were searched.
This signature is the same of the CDF's search for
$FCNC$ decay of the scalar top and scalar bottom
quark: $\tilde{t}_{1} \rightarrow c{\tilde{\chi}}^{0}_{1}$,
$\tilde{b}_{1} \rightarrow c{\tilde{\chi}}^{0}_{1}$.\\
\noindent
Data have been selected using online triggers that requires
large missing-$E_{T}$ ($\not\!\!\!E_{\rm T}$$>35$ $GeV$).
The QCD background is reduced by requiring
$2$ or $3$ jets with $E_{T}>15$ $GeV$ with $|\eta|< 2$ and by excluding
events containing jets with $7$ $GeV<E_{T}<15$ $GeV$ in the region
$|\eta|< 3.6$. Further cuts have been applied in order to remove
fake sources of $\not\!\!\!E_{\rm T}$:
$\Delta\phi(\not\!\!\!E_{\rm T},jet_{1,2})> 45°$,
$\Delta\phi(\not\!\!\!E_{\rm T},jet_{1})< 165°$
and $\Delta\phi(jet_{1},jet_{2})> 165°$
and $W$ and $Z^{0}$are reduced by vetoing events with
high-$p_{T}$ leptons ($e$, $\mu$).
Heavy flavour jets are then identified using a method known as
jet probability
tag~\cite{stopRegina} that calculate, using the SVX informations,
the probability that a cluster tracks form a jet originated from
the primary vertex.
In the $c \bar{c}\nu_{\tau}\bar{\nu}_{\tau}$ channel,
$11$ events are observed with an expected
SM contribution of $14.5 \pm 4.2$;
in the $b \bar{b}\nu_{\tau}\bar{\nu}_{\tau}$ $5$ events are
observed with an expected number of SM events of
$5.8 \pm 1.8$.
Figures~\ref{Leptoquark1} and~\ref{Leptoquark2}
show respectively the limits on $\pi_{LQ}$ production via
color-octet technirho decays and the continuum leptoquark
limits for $LQ2$ and $LQ3$.

\section{Conclusions}
Tevatron experiments performed extensive searches for
physics beyond the Standard Model.
No positive results have been found so far
showing that the data are consistent with
the SM expectations.
CDF and D$\not$0 continue the analysis of Run I data
placing limits on new physics, including Supersymmetry, 
large space-time dimensions and leptoquark models. 
With the Run II upgrades, providing an higher
acceptance and higher luminosity, it will be possible
to make important progresses

\newpage

\vspace*{-0.4cm}
\noindent
in the search for  new phenomena as well as in 
setting limits on a larger variety of theoretical models.

\vspace*{0.4cm}
\section{Acknowledgments}

I would like to express my appreciation to the
organizers and to the other speakers for the
immensely stimulating and enjoyable conference.



\eject


\section*{References}
\begin{thebibliography}{99}
\bibitem{SM}C.~Yang,
  {\it  In Beijing 1998, Physics and detector at the linear collider 37-56}.

\bibitem{Altarelli}
   G.~Altarelli, CERN-TH-98-348, {\bf hep-ph/9811456} (1998).

\bibitem{SUSYTEO}
   S.P.~Martin, {\bf hep-ph/9709356}.

\bibitem{BOER}
   W.~de Boer, {\bf hep-ph/9808448}

\bibitem{DREINER}
   H.~Dreiner, {\it Pramana} {\bf 51}, 123 (1998).

\bibitem{VISSANI}
  J.~Ellis {\it et al.}, {\PLB}{150}{142}{1985}.\\
  F.~Vissani,  IC-96-32, {\bf hep-ph/9602395} (1995).

\bibitem{GENSTOP}
  Ken-ichi~Hikasa, Makoto~Kobayashi,
   \Journal{\PRD}{36}{724}{1987}.

\bibitem{stop1}
  C.~Holck, hep-ex/9903060.\\
  C.~Pagliarone,
     {\it Proceedings of 13th Topical Conference on Hadron Collider
     Physics, Mumbai, India, 14-20 Jan 1999}, FERMILAB-CONF-99-062-E.

\bibitem{cdf_cord}
  In the CDF and  D$\not$O coordinate system, $\phi$ is the azimuthal angle and
  $\theta $ is the polar angle with respect
  to the proton beam direction. The pseudorapidity $\eta $ is defined as
  $\eta=- \ln \tan (\theta / 2)$. The transverse momentum of a particle is
  $p_{\rm T} = p \sin \theta $. If the magnitude of this vector is obtained
  using the calorimeter energy rather than the spectrometer momentum, it
  becomes the transverse energy $E_{\rm T}$.
  Jets are defined as clusters of energy in $\eta-\phi $ space with a
  fix cone size.
  The missing transverse energy ($\not\!\!\!E_{\rm T}$) is defined
  as the difference between the vector sum of all the transverse
  energies and zero.

\bibitem{cdf_top}
  F. Abe {\it et al.}, \ Phys. Rev. Lett. {\bf 74}, 2626 (1995).


\bibitem{gluinoclassic}
   F.    Abe {\it et al.},\ Phys.\ Rev.\ Lett.\ {\bf 69}, 3439 (1992);\\
   S. Abachi {\it et al.},\ Phys.\ Rev.\ Lett.\ {\bf 75}, 618 (1995).


%

\bibitem{RPVbiblio}
   B.~Abbott {\it et al.},\ Phys.\ Rev.\ Lett.\  {\bf 80} 1591 (1998).

\bibitem{Arkani1}
   N.~Arkani-Hamed, S.~Dimopoulos and G.~Dvali,
   Phys.\ Lett.\  {\bf B429}, 263 (1998);\\
   I.~Antoniadis, N.~Arkani-Hamed, S.~Dimopoulos and G.~Dvali,
   Phys.\ Lett.\  {\bf B436}, 257 (1998);

\bibitem{Arkani2}
   N.~Arkani-Hamed, S.~Dimopoulos and G.~Dvali,
   Phys.\ Rev.\  {\bf D59} (1999)

\bibitem{leptoquarks}
   J.C.~Pati and A.~Salam, Phys.\ Rev.\ {\bf D10}, 274 (1974);\\
   H.~Georgi and S.L.~Glashow, Phys.\ Rev.\ Lett.\ {\bf 32}, 438 (1974);\\
   J.L.~Hawett and T.G.~Rizzo,  Phys.\ Rev.\ Lett.\ {\bf 183}, 193 (1989).

\bibitem{acosta}
   D.E.~Acosta and S.K.~Blessing, Annu.\ Rev.\ Nucl.\ Sci.\ {\bf 49} 389 (1999);\\
   F.~Abe {\it et al.}, Phys.\ Rev.\ Lett.\  {\bf 79}, 4327 (1997);\\
   B.~Abbott {\it et al.}, Phys.\ Rev.\ Lett.\  {\bf 84} 2088 (2000);\\
   B.~Abbott {\it et al.}, Phys.\ Rev.\ Lett.\  {\bf 83} 2896 (1999);\\
   B.~Abbott {\it et al.}, Phys.\ Rev.\ Lett.\  {\bf 80} 2051 (1998);\\
   B.~Abbott {\it et al.}, Phys.\ Rev.\ Lett.\  {\bf 79} 4321 (1997).

\bibitem{stopRegina}
  T.~Affolder {\it et al.}, \ Phys.\ Rev.\ Lett.\  {\bf 84}, 5273 (2000). 

\end{thebibliography}
\end{document}